\documentclass{ws-procs9x6}

\usepackage{subfigure}

\begin{document}

\title{Technicolor in the LHC Era\footnote{This talk from KMIIN 2011 reports on work first published by the authors as Ref. 1, and
a more detailed discussion of these results can be found there.}}

\author{R.\ Sekhar Chivukula\footnote{Speaker at conference}
, Pawin Ittisamai, Jing Ren\footnote{Also at {\it Center for High Energy Physics and Institute of Modern Physics, 
Tsinghua University, 
Beijing 100084, China.}} , and Elizabeth H.\ Simmons}

\address{Department of Physics and Astronomy,\\
Michigan State University,\\
East Lansing, MI 48824, USA\\
E-mail: sekhar@msu.edu, ittisama@msu.edu, jingren@pa.msu.edu, \\
and esimmons@pa.msu.edu}

\begin{abstract}
LHC searches for the standard model Higgs Boson in $\gamma\gamma$ or $\tau\tau$  decay modes
place strong constraints on the light technipion state predicted in technicolor models that include colored technifermions.
Compared with the standard Higgs Boson, the technipions have an enhanced production rate (largely because the technipion decay constant
is smaller than the weak scale) and also enhanced branching ratios into di-photon and di-tau final states (largely due to the suppression of
$WW$ decays of the technipions).  Recent ATLAS and CMS searches for Higgs bosons exclude the presence of technipions with masses from 110 GeV to nearly $2 m_t$ in technicolor models that (a) include colored technifermions (b) feature topcolor dynamics and (c) have technicolor groups with three or more technicolors ($N_{TC} \geq 3$).  \end{abstract}


\bodymatter

\section{Introduction}

Experiments now underway at the Large Hadron Collider are striving to discover the agent of electroweak symmetry breaking, thereby revealing the origin of the masses of the elementary particles.   Many of the searches are phrased in terms of placing constraints
on the properties of the scalar Higgs boson ($h_{SM}$) predicted to exist in the standard model \cite{standard-model-a, standard-model-b, standard-model-c}.  
Recently, both the ATLAS  and CMS collaborations at the CERN LHC have reported searches for the standard model Higgs in the two-photon \cite{Collaboration:2011ww, CMS-higgs-di-gamma}
 and $\tau^+ \tau^-$ \cite{ATLAS-higgs-di-tau-low, ATLAS-higgs-di-tau-high, CMS-higgs-di-tau} decay channels.  They have placed upper bounds on the cross-section times branching ratio ($\sigma \cdot B$)  in each channel over the approximate mass range 110 GeV $\leq m_{h} \leq$ 145 GeV, generally finding that $\sigma\cdot B$ cannot exceed the standard model prediction by more than a factor of a few.  In addition, ATLAS has independently constrained the production of a heavy neutral scalar SM Higgs boson with mass up to 600 GeV and decaying to $\tau^+ \tau^-$.  In this paper we apply these limits to the neutral ``technipion'' ($\Pi_T$) states predicted to exist in technicolor models that include colored technifermions.  Because both the technipion production rates and their branching fractions to $\gamma\gamma$ or $\tau\tau$ can greatly exceed the values for a standard model Higgs, the LHC results place strong constraints on technicolor models.  This strategy was first suggested for hadron supercolliders over fifteen years ago in Refs. \refcite{Eichten:1984eu,Eichten:1986eq,Chivukula:1995dt}.
 
Technicolor \cite{Weinberg:1975gm, Weinberg:1979bn,Susskind:1978ms} is a dynamical theory of electroweak symmetry breaking in which a new strongly-coupled gauge group (technicolor) causes bilinears of the fermions carrying its gauge charge (technifermions) to acquire a non-zero vacuum expectation value.  If the technifermion bilinear carries appropriate weak and hypercharge values, the vacuum expectation value breaks the electroweak symmetry to its electromagnetic subgroup.  Fermion masses can then be produced dynamically if technicolor is incorporated into a larger ``extended technicolor''   \cite{Dimopoulos:1979es, Eichten:1979ah} framework coupling technifermions to the ordinary quarks and leptons.   Producing realistic values of fermion masses from extended technicolor (ETC) interactions without simultaneously generating large flavor-changing neutral currents (FCNC) is difficult; the best prospects are ``walking'' technicolor models where the presence of many technifermion flavors causes the technicolor gauge coupling to vary only slowly with energy scale \cite{Holdom:1981rm,Holdom:1984sk,Yamawaki:1985zg,Appelquist:1986an,Appelquist:1986tr,Appelquist:1987fc}.  Even in those models, it is difficult to generate the observed mass of the top quark from ETC interactions without producing unacceptably large weak isospin violation \cite{Chivukula:1988qr}; the best known solution is to generate most of the top quark's mass via new strong ``topcolor''   \cite{Hill:1991(a)t} dynamics, without a large contribution from ETC \cite{Hill:1994hp}.  

Many technicolor models, \cite{Hill:2002ap} including those with walking and topcolor dynamics, feature technipion states, pseudo-scalar bosons that are remnants of electroweak symmetry breaking in models with more than one weak doublet of technifermions.   
It has been shown\cite{Belyaev:2005ct} that technipions can be produced at a greater rate than the standard model Higgs at hadron colliders, because the technipion decay constant is smaller than the electroweak scale, and also that the technipions can have higher branching fractions to $\gamma\gamma$ or $\tau\tau$ final states.  As a result, the technipions are predicted to produce larger signals in these two channels at LHC than the $h_{SM}$ would \cite{Belyaev:2005ct}.   

In this work, we show that the ATLAS \cite{Collaboration:2011ww, ATLAS-higgs-di-tau-low, ATLAS-higgs-di-tau-high} and CMS \cite{CMS-higgs-di-gamma, CMS-higgs-di-tau} searches for the standard model Higgs exclude, at 95\% CL, technipions of masses from 110 GeV to nearly $2 m_t$ in technicolor models that (a) include colored technifermions (b) feature topcolor dynamics and (c) have technicolor groups with three or more technicolors ($N_{TC} \geq 3$).   For certain models of this kind, the limits also apply out to higher technipion masses or down to the minimum number of technicolors  ($N_{TC} = 2$).  We also illustrate how the limits may be modified in models in which extended technicolor plays a significant role in producing the mass of the top quark; in some cases, this makes little difference, while in other cases the limit is softened somewhat.  Overall, we find that ATLAS and CMS significantly constrain technicolor models.  Moreover, as the LHC collaborations collect additional data on these di-tau and di-photon final states and extend the di-photon analyses to higher mass ranges, they should be able to quickly expand their reach in technicolor parameter space.

\section{Technicolor and Technipions}
\label{asec:technicp}



Many dynamical symmetry-breaking models \cite{Hill:2002ap}  include more than the minimal two flavors of technifermions needed to break the electroweak symmetry.  In that case, there will exist light pseudo Nambu-Goldstone bosons known as technipions, which could potentially be accessible to a standard Higgs search. Technipions that are bound states of colored technifermions can be produced through quark or gluon scattering at a hadron collider, like the LHC, through the diagrams in Figure 1.   In the models with topcolor dynamics, where ETC interactions (represented by the shaded circle) contribute no more than a few GeV to the mass of any quark, there is only a small ETC-mediated coupling between the technipion and ordinary quarks in diagrams 1(b) and 1(c).  Combining that information with the large size of the gluon parton distribution function (PDF)  at the LHC and the $N_{TC}$ enhancement factor in the techniquark loop at left, we expect that the diagram in Figure 1(a) will dominate technipion production in these theories, which we study here and in Section \ref{asec:topcol}.  Technipions in models without strong top dynamics could, in contrast, have a large top-technipion coupling, making diagram 1(c) potentially important; we will consider that scenario  briefly\footnote{For a more complete discussion, see \protect\refcite{Chivukula:2011ue}.} at the end of Section \ref{asec:topcol}.  

	\begin{figure}[t,b]
	\centering
	\subfigure{
		\includegraphics[scale=0.15]{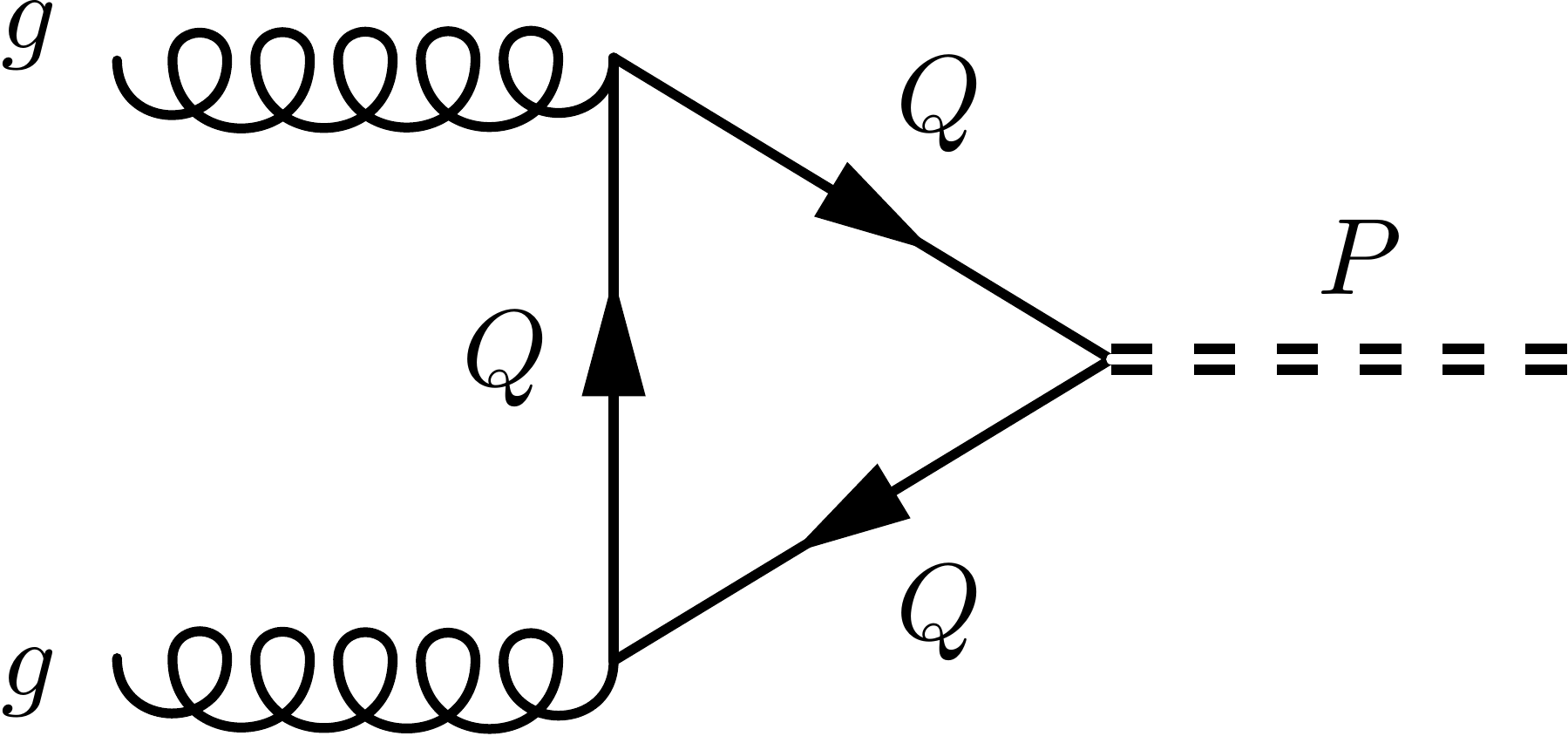}
	} \qquad
	\subfigure{
		\includegraphics[scale=0.15]{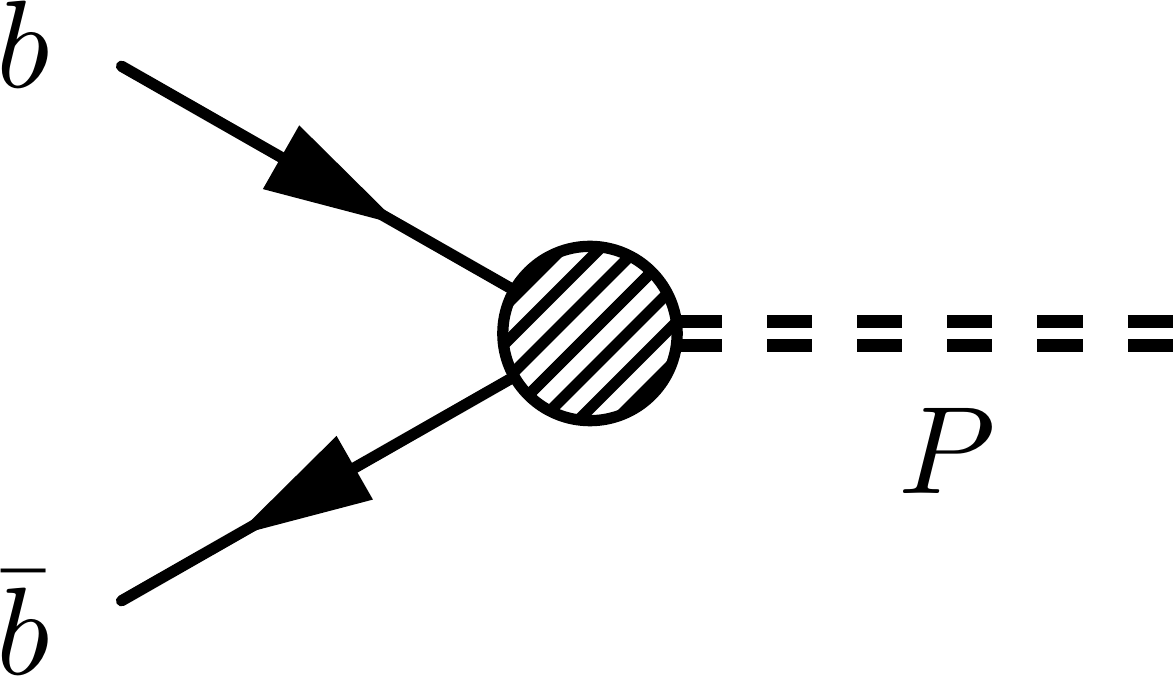}
	} \qquad
	\subfigure{
		\includegraphics[scale=0.15]{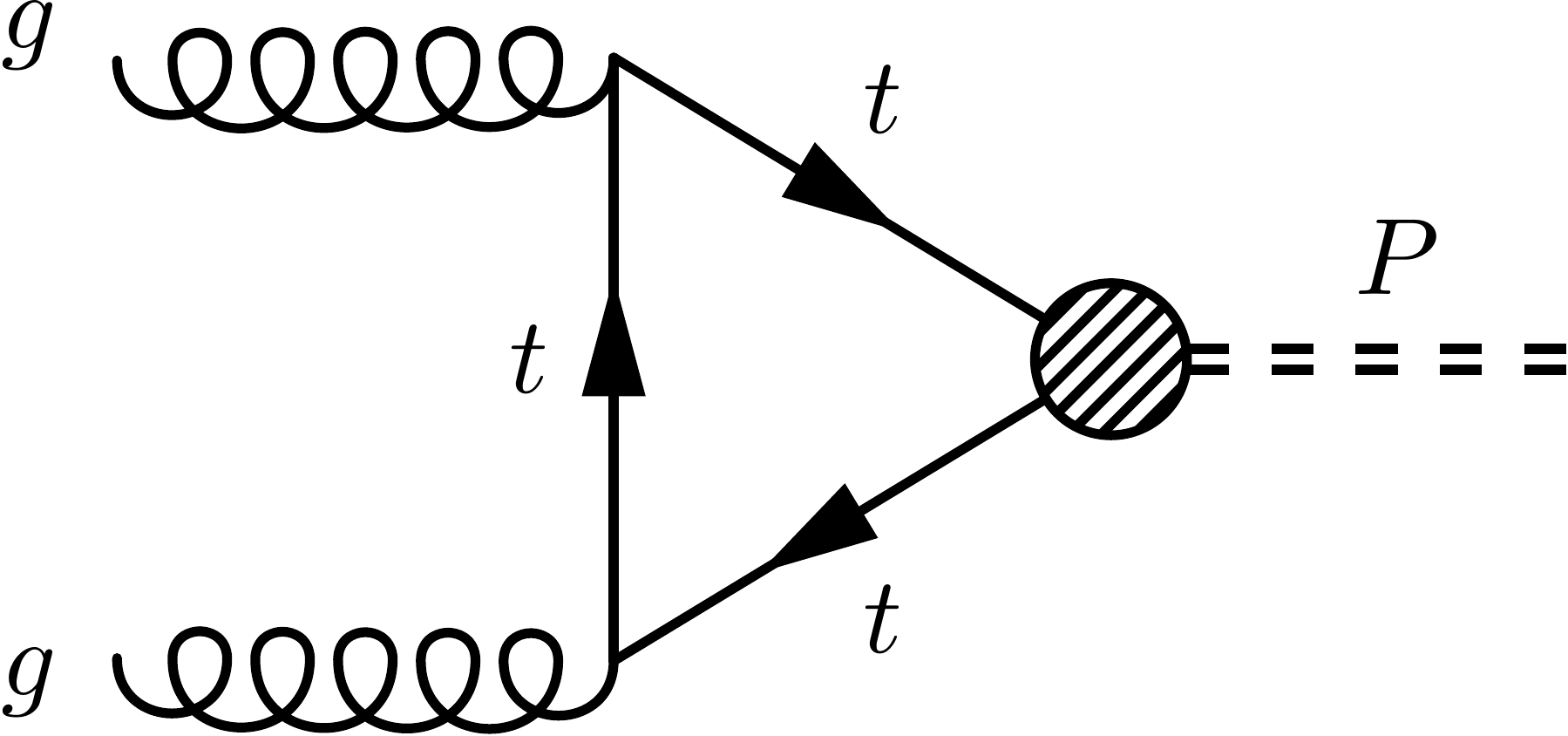}
	}
	\caption{Feynman diagrams for single technipion production through gluon fusion through a loop of colored technifermions, $b\bar{b}$ annihilation,  and gluon fusion
	through a top-quark loop at LHC.  The shaded circles represent an ETC coupling between the ordinary quarks and techniquarks.}
	\label{fig:feynman-diagrams}
	\end{figure}

No single technicolor model has been singled out as a benchmark; rather, different classes of models have been proposed to address the challenges of dynamically generating mass while complying with precision electroweak and flavor constraints.   We will study the general constraints that the current LHC data can place on a variety of theories with colored technifermions and light technipions.  Following \cite{Belyaev:2005ct}, the specific models we examine are: 1) the original one-family model of Farhi and Susskind \cite{Farhi:1980xs} with a full family of techniquarks and technileptons, 2) a variant on the one-family model \cite{Casalbuoni:1998fs} in which the lightest technipion contains only down-type technifermions and is significantly lighter than the other pseudo Nambu-Goldstone
bosons,  3) a multiscale walking technicolor model \cite{Lane:1991qh} designed to reduce flavor-changing neutral currents,
4) a low-scale technciolor model (the Technicolor Straw Man -- TCSM -- model) \cite{Lane:1999uh} with many weak doublets of technifermions and 5) a one-family model with weak-isotriplet technifermions \cite{Manohar:1990eg}.  Properties of the lightest electrically-neutral technipion in each model that couples to gluons (and can therefore be readily produced at LHC) are shown in Table \ref{tab:technipions}.  For completeness, in the figure caption we show the name and technifermion content of each state in the notation of the original paper proposing its existence.
For simplicity, in what follows  the lightest relevant neutral technipion of each model will be generically denoted $P$. Furthermore, we will assume that the lightest technipion state is significantly lighter than other neutral (pseudo)scalar technipions in the spectrum, in order to facilitate the comparison to the standard model Higgs boson.\footnote{ The detailed spectrum of any technicolor model depends on multiple factors, particularly the parameters describing the  ``extended technicolor" \cite{Dimopoulos:1979es,Eichten:1979ah} interaction that transmits electroweak symmetry breaking to the ordinary quarks and leptons.  Models in which several light neutral PNGBs are nearly degenerate could produce even larger signals than those discussed here.}

Single production of a technipion can occur through the axial-vector anomaly which
couples the technipion to pairs of gauge bosons. For an $SU(N_{TC})$ technicolor group
with technipion decay constant $F_P$, the anomalous coupling between the technipion and
a pair of gauge bosons is given, in direct analogy with the coupling of a QCD pion to
photons,\footnote{Note that the normalization used here is identical to that in \cite{Belyaev:2005ct} and differs
from that used in \cite{Lynch:2000hi} by a factor of 4.}
 by \cite{Dimopoulos:1980yf,Ellis:1980hz,Holdom:1981(b)g}
\begin{equation}
N_{TC} {\cal A}_{V_1 V_2} {\frac{g_1 g_2}{8 \pi^2 F_P}} \epsilon_{\mu\nu\lambda\sigma}
k_1^\mu k_2^\nu \epsilon_1^\lambda \epsilon_2^\sigma
\label{eq:anom}
\end{equation}
where 
\begin{equation}
{\cal A}_{V_1 V_2} \equiv Tr \left[ T^a ( T_1 T_2 + T_2 T_1)_L + T^a (T_1 T_2 + T_2 T_1)_R \right]
\end{equation}
is the anomaly factor, $T^a$ is the generator of the axial vector current associated with the techipion, subscripts $L$ and $R$ denote the left- and right-handed technifermion components of the technipion, the $T_i$ and $g_i$ are the generators and couplings associated with gauge bosons $V_i$, and the $k_i$ and $\epsilon_i$ are the four-momenta and polarizations of the gauge
bosons. The value of the anomaly factor ${\cal A}_{gg}$ for the lightest PNGB of each model that is capable of coupling to gluons appears in Table 
\ref{tab:technipions}, along with the anomaly factor ${\cal A}_{\gamma\gamma}$ coupling the PNGB to photons.  Also shown in the table is the value of the technipion decay constant, $F_P$ for each model.\footnote{ In the multi-scale model [model 3], various technicondensates form at different scales; we set $F_P^{(3)}$ = $ \frac{v}{4}$ in keeping with \cite{Lane:1991qh}  and to ensure that the technipion mass will be in the range to which the standard Higgs searches are sensitive. }

Examining the technipion wavefunctions in Table \ref{tab:technipions} we note that the PNGB's do {\it not} decay to $W$ boson pairs, since the $W^+W^-$ analog of Figure 1(a)  vanishes due to a cancellation between techniquarks and technileptons.  The corresponding $ZZ$ diagrams will not vanish but, again due to a cancellation between techniquarks and technileptons, will instead yield small couplings for the technipion to
$ZZ$ (and $Z\gamma$) proportional to the technifermion hypercharge couplings \cite{Lynch:2000hi}. The small coupling and phase space suppression yield much smaller branching ratios for the PNGB's to decay
to $ZZ$ or $Z\gamma$, and hence these modes are irrelevant to our limits.

	\begin{table}[!bt]
	\tbl{Properties of the lightest relevant PNGB (technipion)  in representative technicolor models with colored technifermions.  In each case, we show the name and technifermion content of the state (in the notation of the original paper), the ratio of the weak scale to the technipion decay constant, the anomaly factors for the two-gluon and two-photon couplings of the technipion, and the technipion's  couplings to leptons and quarks.  The symbols``Q" or ``D" refer to color-triplets (a.k.a. techniquarks) while those including ``L" or ``E" refer to color-singlets (a.k.a. technileptons).  The multiscale model incorporates six technileptons, which we denote by $L_{\ell}$. For the TCSM low-scale model, $N_D$ refers to the number of weak-doublet technifermions contributing to electroweak symmetry breaking; this varies with the size of the technicolor group.    The parameter $y$ in the isotriplet model is the hypercharge assigned to the technifermions.}
{\begin{tabular}{@{}c||c|c|c|c|c|c|c@{}}
\hline\hline
TC models & \multicolumn{2}{c|}{PNGB and content} & $v/F_P$ & $A_{gg}$ & $A_{\gamma\gamma}$ & $\lambda_l$ & $\lambda_f$\\
\hline\hline
FS one family\cite{Farhi:1980xs} & $P^1$ & $\frac{1}{4\sqrt{3}}(3\bar{L}\gamma_5L-\bar{Q}\gamma_5Q)$ & 2 & $-\frac{1}{\sqrt{3}}$ & $\frac{4}{3\sqrt{3}}$ & 1 & 1 \\
\hline
Variant one family\cite{Casalbuoni:1998fs} & $P^0$ & $\frac{1}{2\sqrt{6}}(3\bar{E}\gamma_5E-\bar{D}\gamma_5D)$ & 1 & $-\frac{1}{\sqrt{6}}$ & $\frac{16}{3\sqrt{6}}$ & $\sqrt{6}$ & $\sqrt\frac{2}{3}$\\
\hline
LR multiscale\cite{Lane:1991qh} & $P^0$ & $\frac{1}{6\sqrt{2}}(\bar{L}_{\ell}\gamma_5L_{\ell}-2\bar{Q}\gamma_5Q)$ & 4 & $-\frac{2\sqrt{2}}{3}$ & $\frac{8\sqrt{2}}{9}$ & 1 & 1 \\
\hline
TCSM low scale\cite{Lane:1999uh} & $\pi^{0'}_T$ & $\frac{1}{4\sqrt{3}}(3\bar{L}\gamma_5L-\bar{Q}\gamma_5Q)$ & $\sqrt{N_D}$ & $-\frac{1}{\sqrt{3}}$ & $\frac{100}{27\sqrt{3}}$ & 1 & 1 \\
\hline
MR Isotriplet \cite{Manohar:1990eg} & $P^1$ & $\frac{1}{6\sqrt{2}}(3\bar{L}\gamma_5L-\bar{Q}\gamma_5Q)$ & 4 & $-\frac{1}{\sqrt{2}}$ & $24\sqrt{2}y^2$ & 1 & 1  \\
\hline\hline
\end{tabular}}
	\label{tab:technipions}
	\end{table}

The rate of single technipion production via glue-glue fusion and a techniquark loop (Figure 1(a)) is proportional to the technipion's decay width to gluons through that same techniquark loop
\begin {equation}
{\Gamma (P \rightarrow gg)} = { \frac{m_{P}^3}{ 8 \pi}}   \left(\frac {\alpha_s N_{TC}{\cal A}_{gg}}{2 \pi F_P} \right)^2\ .
\label{eq:techni-glu}
\end {equation}
 In the  SM, the equivalent expression (for Higgs decay through a top quark loop) looks like \cite{Gunion:1989we} 
\begin {equation}
{\Gamma (h_{SM} \rightarrow gg)} ={\frac{m_{h}^3}{8 \pi}}  \left(\frac{ \alpha_s}{3 \pi v}\right)^2  
\left[\frac{3\tau}{2}(1 + (1-\tau) f(\tau))\right]^2~,
\label{eq:higgs-tau}
\end {equation}
where $\tau \equiv (4 m_t^2 / m_h^2)$ and
\begin{equation}
f(\tau) = 
\begin{cases} \left[ \sin^{-1} (\tau^{-{\frac12}}) \right]^2& \text{if $\tau \geq 1$}
\\
-\frac14 \left[ \log \left( \frac{1 + \sqrt{1-\tau}}{1 - \sqrt{1-\tau}}\right) - i \pi \right]^2&\text{if $\tau < 1$.}
\end{cases}
\label{eq:fftau}
\end{equation}
so that the expression in square brackets in Eq. (\ref{eq:higgs-tau}) approaches 1 in the limit where the top quark is heavy ($\tau >> 1$).  Therefore, the rate at which $P$ is produced from $gg$ fusion exceeds that for a standard Higgs of the same mass by a factor
\begin {equation}
\kappa_{gg\ prod} = \frac{ \Gamma (P \rightarrow gg)}{ \Gamma (h_{SM} \rightarrow gg)} = 
 \frac{9}{4} N_{TC}^2 {\cal A}_{gg}^2 \frac{v^2}{F_P^2} 
\left[\frac{3\tau}{2}(1 + (1-\tau) f(\tau))\right]^{-2}
\label{eq:kappagg}
\end {equation}
 where, again, the factor in square brackets is 1 for scalars much lighter than $2m_t$.  A large technicolor group and a small technipion decay constant can produce a significant enhancement factor.

Technipions can also be produced at hadron colliders via $b\bar{b}$ annihilation (as in Figure 1(b)), because the 
 ETC interactions coupling quarks to techniquarks afford the technipion a decay mode into fermion/anti-fermion pairs.  
The rate is proportional to the technipion decay width into fermions:
\begin {equation}
{ \Gamma (P \rightarrow f \overline{f})} = {\frac{N_C\, \lambda^2_f\, m^2_f\, m_P}{8 \pi\, F^2_P}}\,
\left(1 -  \frac{4m_f^2}{m_P^2}\right)^{\frac{s}{2}}
\label{eq:phasesp}
\end {equation}
where  $N_C$ is 3 for quarks and 1 for leptons.  The phase space exponent,
$s$, is 3 for scalars and 1 for pseudoscalars; the lightest PNGB in our technicolor models is a
pseudoscalar.  For the technipion masses considered here, the value of
the phase space factor in (\ref{eq:phasesp}) is so close to one that the
value of $s$ makes no practical difference.   The factors $\lambda_f$ are 
non-standard Yukawa couplings distinguishing leptons from quarks.  The variant one-family model has $\lambda_{quark} = \sqrt{\frac{2}{3}}$ and $\lambda_{lepton} =
\sqrt{6}$; the multiscale model also includes a similar factor, but with average
value 1; $\lambda_f=1$ in the other models.   For comparison, the decay width of the SM Higgs into b-quarks is:
\begin {equation}
{ \Gamma (h_{SM} \rightarrow b \overline{b})} = {\frac{3\,m^2_b\,m_h}{8\pi\,v^2}}
\left(1 - \frac {4m_b^2}{m_h^2}\right)^{\frac{3}{2}}
\end {equation}
 Thus, the rate at which $P$ is produced from $b\bar{b}$ annihilation exceeds that for a standard Higgs of the same mass by 
\begin {equation}
\kappa_{bb\ prod} = \frac{ \Gamma (P \rightarrow b \overline{b})}
{ \Gamma (h_{SM} \rightarrow b \overline{b})} = {\frac{\lambda^2_b\, v^2}{ F^2_P}} 
\left(1 - \frac{4m_b^2}{m_h^2}\right)^{\frac{s-3}{2}}
\label{eq:kappabb}
\end {equation}
The enhancement is smaller than that in Eq. (\ref{eq:kappagg}) because there is no loop-derived factor of $N_{TC}$.

\begin{table}[!tb]
\label{tab:BR130}
\tbl{Branching ratios for phenomenologically important modes (in percent) for technipions of mass 130 GeV for $N_{TC}=2,4$ and for a standard model Higgs  \protect\cite{Dittmaier:2011ti} of the same mass.}
{\begin{tabular}{@{}|c|| c|c|| c|c|| c|c|| c|c|| c|c| ||c|@{}}
\hline
  &\multicolumn{2}{c||}{One} &\multicolumn{2}{c||}{Variant} &\multicolumn{2}{c||}{Multiscale} &\multicolumn{2}{c||}{TCSM} &\multicolumn{2}{c|||}{Isotriplet} &\multicolumn{1}{c|}{}\\
  Decay &\multicolumn{2}{c||}{Family} &\multicolumn{2}{c||}{one family} &\multicolumn{2}{c||}{} &\multicolumn{2}{c||}{low-scale} &\multicolumn{2}{c|||}{} &\multicolumn{1}{c|}{SM}\\
\cline{2-11}
Channel 				 &$N_{TC}$	&$N_{TC}$ 		&$N_{TC}$	&$N_{TC}$		&$N_{TC}$	&$N_{TC}$		&$N_{TC}$	&$N_{TC}$	&$N_{TC}$	&$N_{TC}$ 		& Higgs\\
					&=2 & =4	 	&=2 & =4	 	&=2 & =4	 	&=2 & =4	 		&=2 & =4	 			&  \\
\hline\hline
$b\bar{b}$	 		& 77 & 56		&61 & 50 	 	& 64& 36 		& 77 & 56 		 	&60& 31 			& 49\\
$c\bar{c}$	 		& 7 & 5.1		& 0 & 0	 		& 5.8 & 3.2		& 7 & 5.1	 		&5.4& 2.8			& 2.3\\
$\tau^+\tau^-$	 	& 4.5 & 3.3		&32	& 26	 	& 3.8& 2.1		& 4.5 & 3.3		 	&3.5& 1.8 			& 5.5 \\
$gg$				& 12 & 35		& 7& 23	 		& 26 & 59 		& 12 & 35  			&14& 29				& 7.9\\
$\gamma\gamma$	& 0.011 & 0.033	&0.11 & 0.35	& 0.025& 0.056	& 0.088 & 0.26	 	&17& 36				& 0.23\\
$W^+W^-$			& 0 & 0			& 0 & 0	 		& 0 & 0			& 0 & 0	 			&0 & 0 				& 31 \\
\hline
\end{tabular}}
\end{table}

\begin{table}[!bt]
\label{tab:BR350}
\tbl{Branching ratios for phenomenologically important modes (in percent) for technipions of mass 350 GeV for $N_{TC}=2,4$ and for a standard model Higgs  \protect\cite{Dittmaier:2011ti} of the same mass.}
{\begin{tabular}{|c|| c|c|| c|c|| c|c|| c|c|| c|c| ||c|}
\hline
  &\multicolumn{2}{c||}{One} &\multicolumn{2}{c||}{Variant} &\multicolumn{2}{c||}{Multiscale} &\multicolumn{2}{c||}{TCSM} &\multicolumn{2}{c|||}{Isotriplet} &\multicolumn{1}{c|}{}\\
  Decay &\multicolumn{2}{c||}{Family} &\multicolumn{2}{c||}{one family} &\multicolumn{2}{c||}{} &\multicolumn{2}{c||}{low-scale} &\multicolumn{2}{c|||}{} &\multicolumn{1}{c|}{SM}\\
\cline{2-11}
Channel 				 &$N_{TC}$	&$N_{TC}$ 		&$N_{TC}$	&$N_{TC}$		&$N_{TC}$	&$N_{TC}$		&$N_{TC}$	&$N_{TC}$	&$N_{TC}$	&$N_{TC}$ 		& Higgs\\
					&=2 & =4	 	&=2 & =4	 	&=2 & =4	 	&=2 & =4	 		&=2 & =4	 			&  \\
\hline\hline
$b\bar{b}$	 		& 44 & 18		&42 & 20 	 	& 24 & 7.7 		& 44 & 18 		 	&20& 6.2 			& 0.036\\
$c\bar{c}$	 		& 4 & 1.6		& 0 & 0	 		& 2.2 & 0.69		& 4 & 1.6	 		&1.8& 0.56			& 0.0017\\
$\tau^+\tau^-$	 	& 2.6 & 1		&22	& 11	 	& 1.4 & 0.45		& 2.6 & 1		 	&1.2& 0.36 			& 0.0048\\
$gg$				& 49 & 79		& 35& 68 		& 72 & 91 		& 49 & 79  			&34& 41				& 0.085\\
$\gamma\gamma$	& 0.047 & 0.076	&0.54 & 1		& 0.069& 0.087	& 0.36 & 0.58 		&42& 51				& $\sim 0$\\
$W^+W^-$			& 0 & 0			& 0 & 0	 		& 0 & 0			& 0 & 0	 			&0 & 0 				& 68 \\
\hline
\end{tabular}}
\end{table}

For completeness, we note that the branching fraction for a technipion into a photon pair via a techniquark loop is:
\begin {equation}
{\Gamma (P \rightarrow \gamma\gamma)} = { \frac{m_{P}^3}{ 64 \pi}}   \left(\frac {\alpha_s N_{TC}{\cal A}_{\gamma \gamma}}{2 \pi F_P} \right)^2\ .
\label{eq:techni-gam}
\end {equation}
as compared with the result for the standard model Higgs boson (through a top quark loop) \cite{Gunion:1989we} 
\begin {equation}
{\Gamma (h_{SM} \rightarrow \gamma\gamma)} ={\frac{m_{h}^3}{9 \pi}}  \left(\frac{ \alpha}{3 \pi v}\right)^2  
\left[\frac{3\tau}{2}(1 + (1-\tau) f(\tau))\right]^2~,
\label{eq:higgs-gam-gam-tau}
\end {equation}

From these decay widths, we can now calculate the technipion branching ratios to all of the significant two-body final states, taking $N_{TC} = 2$ and $N_{TC} = 4$ by way of example.  In the TCSM low-scale model we set $N_D = 5\ (10)$ for $N_{TC} = 2\ (4)$ to make the technicolor coupling walk; in the Isotriplet model, we set the technifermion hypercharge to the value $y=1$.  We find that the branching ratio values are nearly independent of the size of $M_P$ within the range 110 GeV - 145 GeV and also show little variation once $M_P > 2 m_t$; to give a sense of the patterns, the branching
fractions for $M_P = 130$ GeV are shown in Table 2 and those for $M_P = 350$ GeV are shown in Table 3.  The branching ratios for the SM Higgs at NLO are given
for comparison; these were obtained from the Handbook of LHC Higgs Cross Sections \cite{Dittmaier:2011ti}. The primary differences are the absence of a $WW$ decay for technipions and the enhancement of the two-gluon coupling (implying increased $gg \to P$ production); the di-photon and di-tau decay widths  can also vary moderately from the standard model values.

Pulling this information together, and noting that the PNGBs are narrow resonances, we may define an enhancement factor for the full production-and-decay process $yy \to {P} \to xx$ as the ratio of the products of the width of the (exclusive)
production mechanism  and the branching ratio for the decay: 
\begin{equation}
\kappa_{yy/xx}^{P} = \frac{ \Gamma({P} \to yy) \times BR({P} \to x x)}
         { \Gamma(h_{SM} \to yy) \times BR(h_{SM} \to  x x)} \equiv \kappa_{yy\ prod}\  \kappa_{xx\ decay} \ .
\label{eq:kappa}
\end{equation}
And to include both the gluon fusion and $b$-quark annihilation production channels when looking for a technipion in the specific decay channel ${P} \to xx$, we define
a combined enhancement factor 
\begin{eqnarray}
\kappa_{total/xx}^{P} 
&=&
\frac{\sigma(gg\to{P}\to xx)+\sigma(bb\to{P} \to xx)}
{{\sigma(gg\to h_{SM} \to xx)+\sigma(bb\to h_{SM} \to xx)} }\nonumber\\
&=&
\frac{\kappa_{gg/xx}^{ P}+\sigma(bb\to {P}\to xx)/\sigma(gg\to h_{SM} \to xx)}
{{1+\sigma(bb\to h_{SM} \to xx)/\sigma(gg\to h_{SM} \to xx)} }\nonumber\\
&=&
\frac{\kappa_{gg/xx}^{P} +\kappa_{bb/xx}^{P}\sigma(bb\to h_{SM} \to xx)/\sigma(gg\to h_{SM} \to xx)}
{{1+\sigma(bb\to h_{SM} \to xx)/\sigma(gg\to h_{SM} \to xx)} }\nonumber\\
&\equiv&
[\kappa_{gg/xx}^{P}+\kappa_{bb/xx}^{P} R_{bb:gg}]/
[{1+  R_{bb:gg}}]  .
\label{kappab}
\end{eqnarray}
Here 
 $R_{bb:gg}$ is the ratio of $b\bar{b}$ and $gg$ initiated Higgs boson production in the 
 Standard Model, which can be calculated using the HDECAY program \cite{Djouadi:1997yw}. In practice, as noted in \refcite{Belyaev:2005ct}, the contribution from b-quark annihilation is much smaller than that from gluon fusion for colored technifermions.
 
\section{Models with colored technifermions and a topcolor mechanism}
\label{asec:topcol}

We will now show how the LHC data constrains technipions composed of colored technifermions in theories where the top-quark's mass is generated by new strong
``topcolor"  dynamics \cite{Hill:1994hp} preferentially coupled to third-generation quarks.  In such models, the ETC coupling between ordinary quarks and technifermions (or technipions) is very small, so that gluon fusion through a top-quark loop will be negligible by comparison with gluon fusion through a technifermion loop, as a source of technipion production.

\subsection{LHC Limits on Models with Light Technipions }

Here we report our results for technipions in the 110 - 145 GeV mass range where direct comparison with Higgs production is possible.  We consider final states with pairs of photons or tau leptons, since the LHC experiments have reported limits on the standard model Higgs boson in both channels.

\begin{figure}
\begin{center}
	\subfigure[\ Variant one-family model \cite{Casalbuoni:1998fs}.]{
		\includegraphics[scale=0.40]{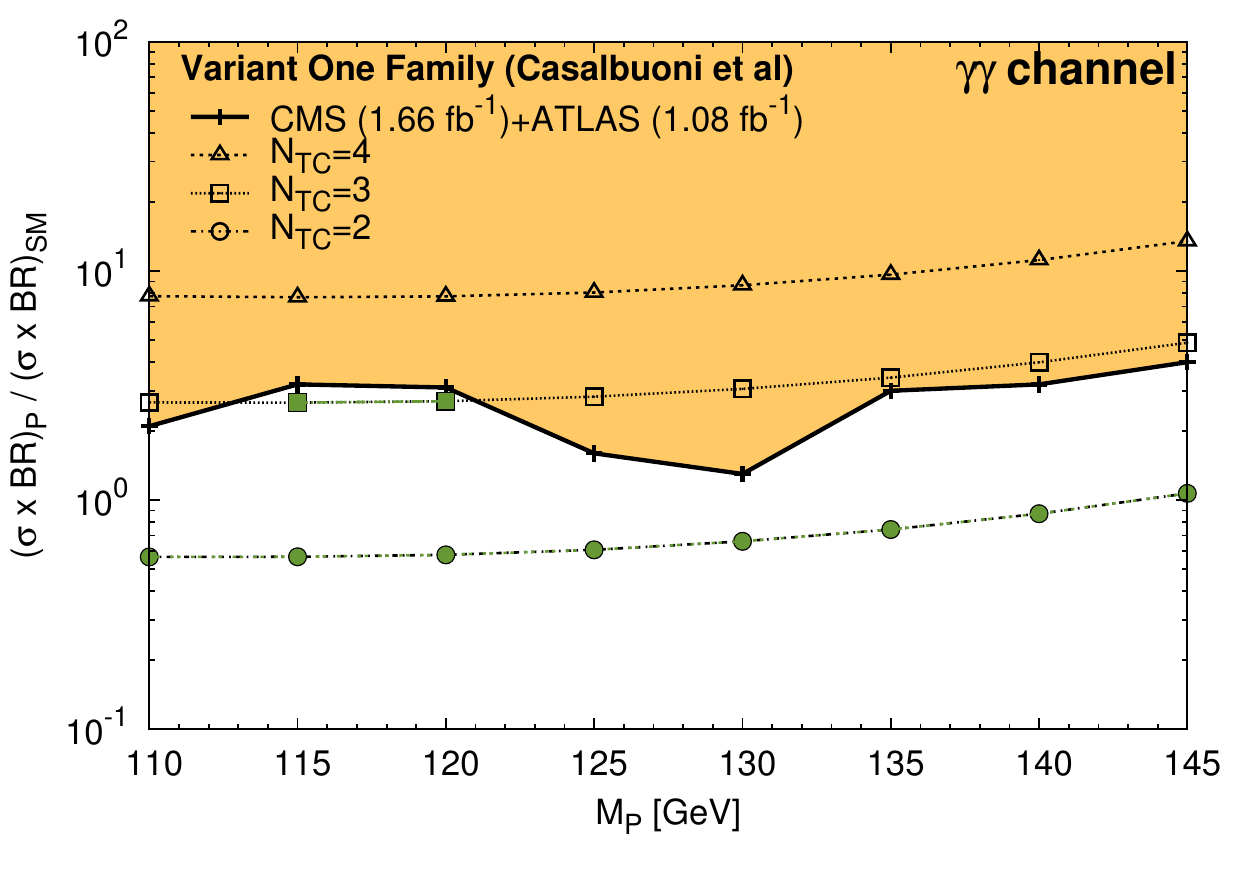}
	}
	\subfigure[\ Isotriplet model  \cite{Manohar:1990eg}. The magnitude of the technifermion hypercharge variable $y$ has been set to 1 for illustration.]{
		\includegraphics[scale=0.40]{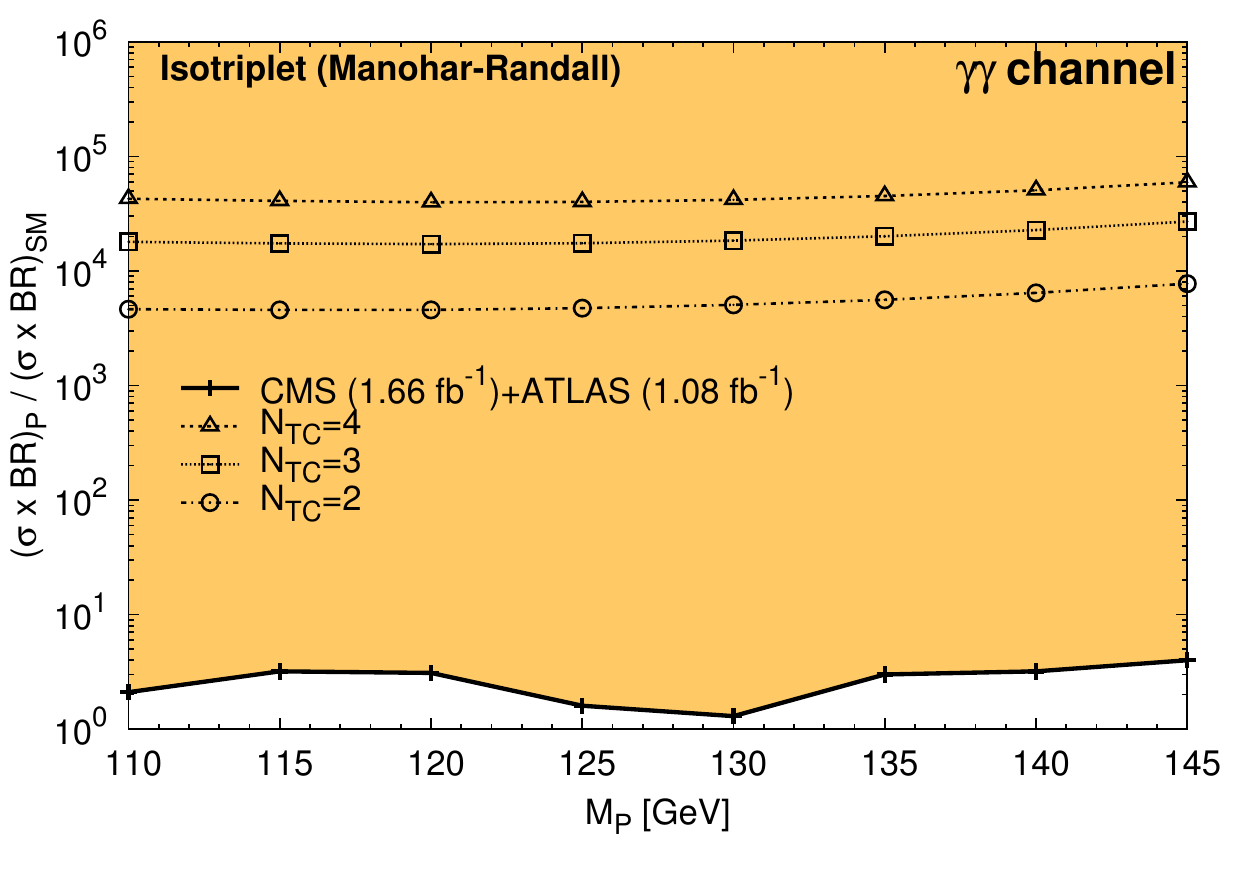}
	}
\end{center}
	\caption{Comparison of experimental limits and technicolor model predictions for production of a new scalar decaying to photon pairs. In each pane, the shaded region (above the solid line) is excluded by the combined $95\%$ CL upper limits on $\sigma_h B_{\gamma\gamma}$ normalized to the SM expectation as observed by CMS \cite{CMS-higgs-di-gamma} and ATLAS \cite{Collaboration:2011ww}.  Each pane also displays (as open symbols)  the theoretical prediction from one of our representative technicolor models with colored technifermions, as a function of technipion mass and for several values of $N_{TC}$.  Values of mass and $N_{TC}$ for a given model that are not excluded by the data are shown as solid (green) symbols.}
	\label{fig:HgamgamEnhancementLimit-Logscale}
	\end{figure}

First, we illustrate the limits derived from the CMS and ATLAS searches for a standard model Higgs boson decaying to $\gamma\gamma$ in two models in Figure \ref{fig:HgamgamEnhancementLimit-Logscale}.  The multiscale \cite{Lane:1991qh}, TCSM low-scale  \cite{Lane:1999uh}, and isotriplet \cite{Manohar:1990eg} models predict rates of technipion production and decay to diphotons that exceed the experimental limits in this mass range even for the smallest possible size of the technicolor gauge group (larger $N_{TC}$ produces a higher rate).   Note that we took the value of the technifermion hypercharge parameter $y$ in the isotriplet model to have the value $y=1$ for purposes of illustration; choosing $y \sim 1/7$ could make this model consistent with the di-photon data for $N_{TC} = 2$, but that would not affect the limits from the di-tau channel discussed below.   For the original \cite{Farhi:1980xs} and variant  \cite{Casalbuoni:1998fs} one-family models, the data still allow $N_{TC} = 2$ over the whole mass range, and $N_{TC} = 3$ is possible for 115 GeV $< M_P <$ 120 GeV; even 135 $< M_P <$ 145 GeV is marginally consistent with the data for $N_{TC}=3$ in the original one-family model.

	\begin{figure}
\begin{center}
	\subfigure[\ Variant one-family model \cite{Casalbuoni:1998fs}.]{
		\includegraphics[scale=0.40]{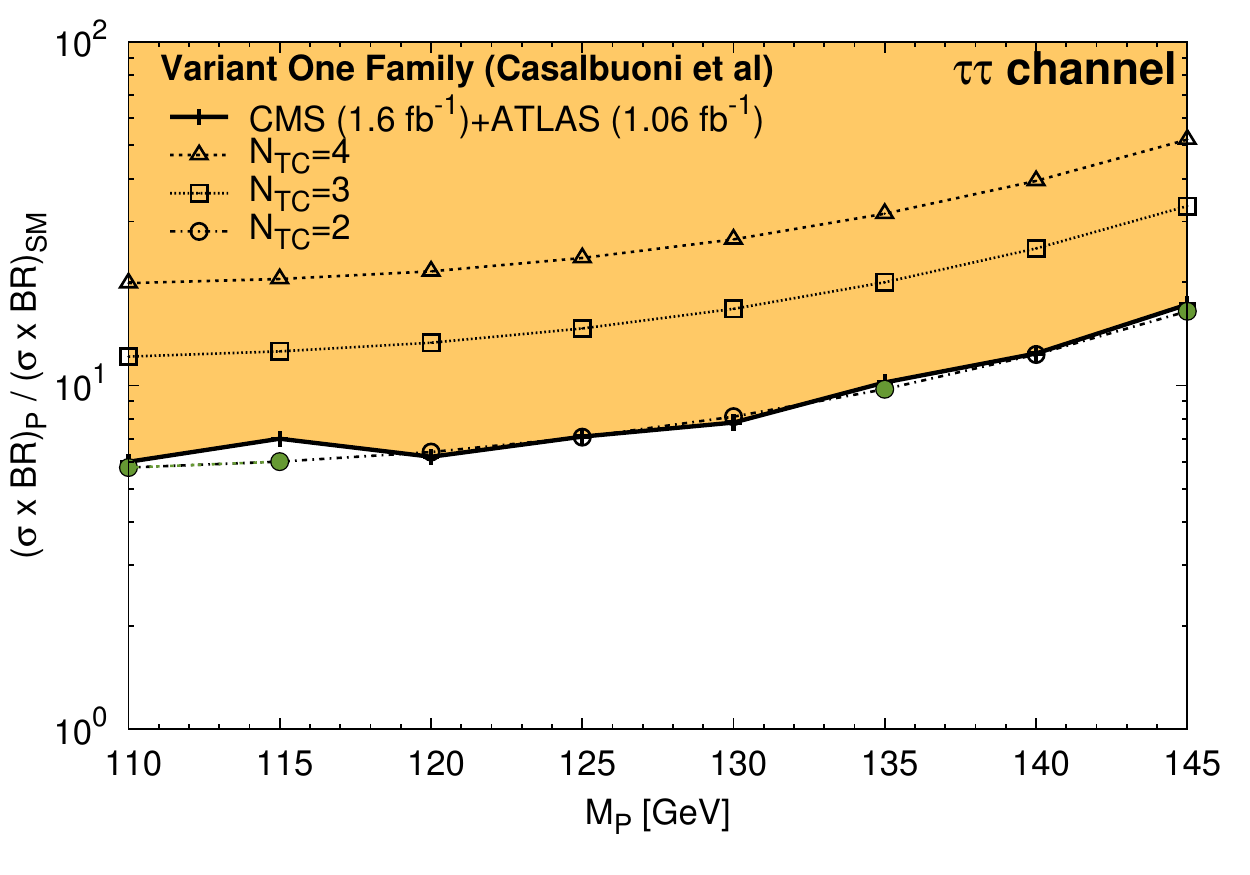}
	}
%
	\subfigure[\ Isotriplet model \cite{Manohar:1990eg}.The magnitude of the technifermion hypercharge variable $y$ has been set to 1 for illustration]{
		\includegraphics[scale=0.40]{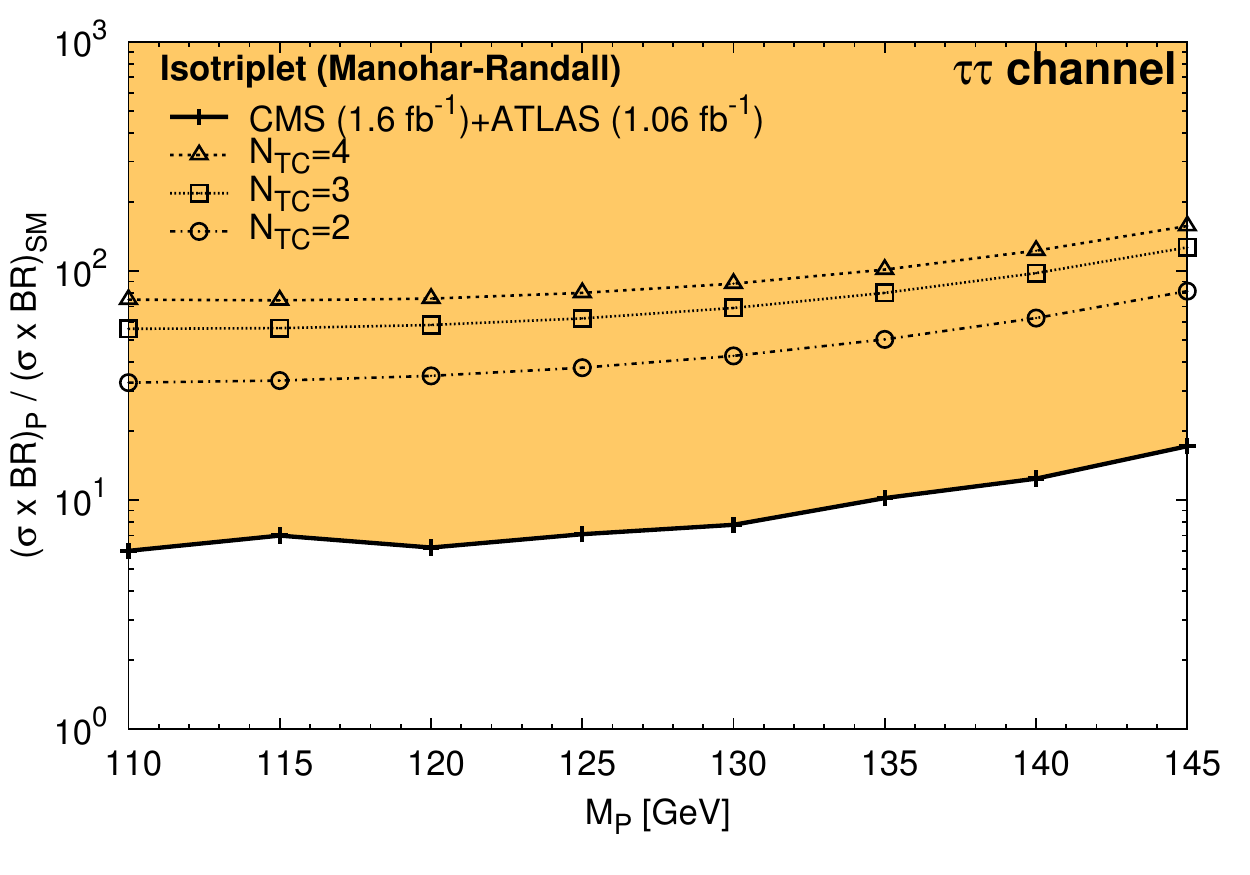}
	}
\end{center}
	\caption{Comparison of experimental limits and technicolor model predictions for production of a new scalar decaying to tau lepton pairs. In each pane, the shaded region (above the solid line) is excluded by the combined $95\%$ CL upper limits on $\sigma_h B_{\tau^+\tau^-}$ normalized to the SM expectation as observed by CMS \cite{CMS-higgs-di-tau} and ATLAS \cite{ATLAS-higgs-di-tau-low}.  Each pane also displays (as open symbols)  the theoretical prediction from one of our representative technicolor models with colored technifermions, as a function of technipion mass and for several values of $N_{TC}$.  Values of $M_P$ and $N_{TC}$ for a given model that are not excluded by the data are shown as solid (green) symbols; the only such point is at $N_{TC}= 2$ and $M_P = 115$ GeV for the variant one-family model.}
	\label{fig:HtautauEnhancementLimit-Logscale}
	\end{figure}

The limits from the CMS and ATLAS searches for a standard model Higgs boson decaying to $\tau^+\tau^-$ in the same mass range are even more stringent, as illustrated in Figure  \ref{fig:HtautauEnhancementLimit-Logscale}.  The data again exclude the multiscale \cite{Lane:1991qh}, TCSM low-scale  \cite{Lane:1999uh}, and isotriplet \cite{Manohar:1990eg} models across the full mass range and for any size of the technicolor gauge group.  The original one-family model\cite{Farhi:1980xs} is likewise excluded; only $M_P = 115$ GeV for $N_{TC} = 2$ is even marginally consistent with data.  The variant  \cite{Casalbuoni:1998fs} one-family model is marginally consistent with data for $N_{TC} = 2$ but excluded for all higher values of $N_{TC}$.  Forthcoming LHC data on $\tau\tau$ final states should provide further insight on these two models for $N_{TC} = 2$.

\subsection{LHC Limits on Heavier Technipions Decaying to Tau-Lepton Pairs}

We now consider technipions that are too heavy to be directly compared with a Higgs in the LHC data, but which can be directly constrained by looking at data from final states with tau-lepton pairs.  ATLAS has obtained \cite{ATLAS-higgs-di-tau-high}  limits on the product of the production cross section with the branching ratio to tau pairs at $95\%$ confidence level for a generic scalar boson in the mass range $100-600$ GeV.   We use this limit to constrain technicolor models as follows.   The production cross section $\sigma(gg\rightarrow P)$ for technicolor models can be estimated by scaling from the standard model\footnote{The standard model production cross section $ \sigma(gg\rightarrow h_{SM})$ at several values of the Higgs mass can be obtained from the Handbook \cite{Dittmaier:2011ti}. } using the production enhancement factor calculated for each technicolor model \cite{Belyaev:2005ct}.  And the branching fraction of the technipions into tau pairs is shown in Table II, above.  Therefore,
	\begin{eqnarray}
	\sigma(gg\rightarrow P){BR(P\rightarrow\tau\tau)} &=& \kappa_{gg \  prod} \sigma(gg\rightarrow h_{SM}) BR(P \rightarrow \tau\tau)\, .
	\end{eqnarray}
Our comparison of the experimental limits with the model predictions is shown in figure \ref{fig:XSBRggHtautauATLASTau-log-lambda-0-5}.  

The data excludes technipions in the mass range from 145 GeV up to nearly $2m_t$ in all models for $N_{TC} \geq 3$.  For the multiscale and isotriplet models, $N_{TC} = 2$ is excluded as well in this mass range; for the TCSM low-scale model, $N_{TC} = 2$ is excluded up to nearly 300 GeV (the few points that are allowed at low mass on this plot are excluded by the data discussed above); while for the original and variant one-family models, $N_{TC} = 2$ can be consistent with data at these higher masses.   Again, further LHC data on di-tau final states will be valuable for discerning whether the models with only two technicolors remain viable.  At present, technicolor models with colored technifermions are strongly constrained even if their lightest technipion is just below the threshold at which it can decay to top-quark pairs.

Moreover, the data also impacts technipions in the mass range above $2 m_t$ in some cases:  $M_P \leq 450$ GeV (375 GeV) is excluded for any size technicolor group in the multiscale (isotriplet) model and $M_P \leq 375$ GeV is excluded for $N_{TC} \geq 3$ in the TCSM low-scale model.  

\subsection{Models with colored technifermions and a top mass generated by ETC}
\label{asec:topetc}

The limits discussed above apply only in cases where the technipion has a very small branching fraction into top quarks, and the branching fraction to di-taus just varies smoothly with the increasing mass of the technipion.   Limits on technipions heavier than $2 m_t$ would not hold in models where extended technicolor dynamically generates the bulk of the top quark mass and the technipion has an appreciable top-quark branching fraction.  In such models, the ETC coupling between the top quark and technipion can be relatively large, which has several consequences.

\begin{figure}
\begin{center}
	\subfigure[\ Original one-family model \cite{Farhi:1980xs}.]{
		\includegraphics[scale=0.40]{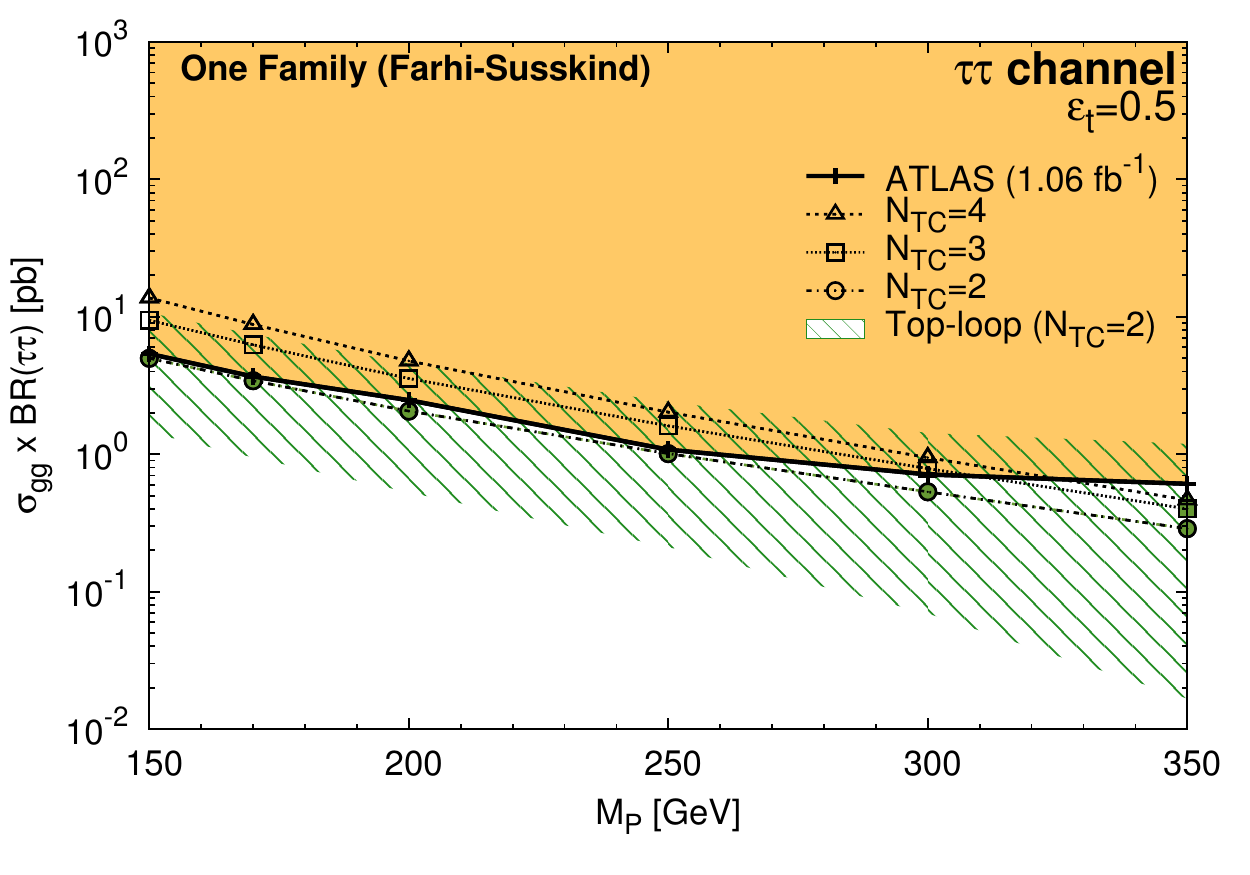}
	}
	\subfigure[\ Multiscale walking technicolor model \cite{Lane:1991qh}.]{
		\includegraphics[scale=0.40]{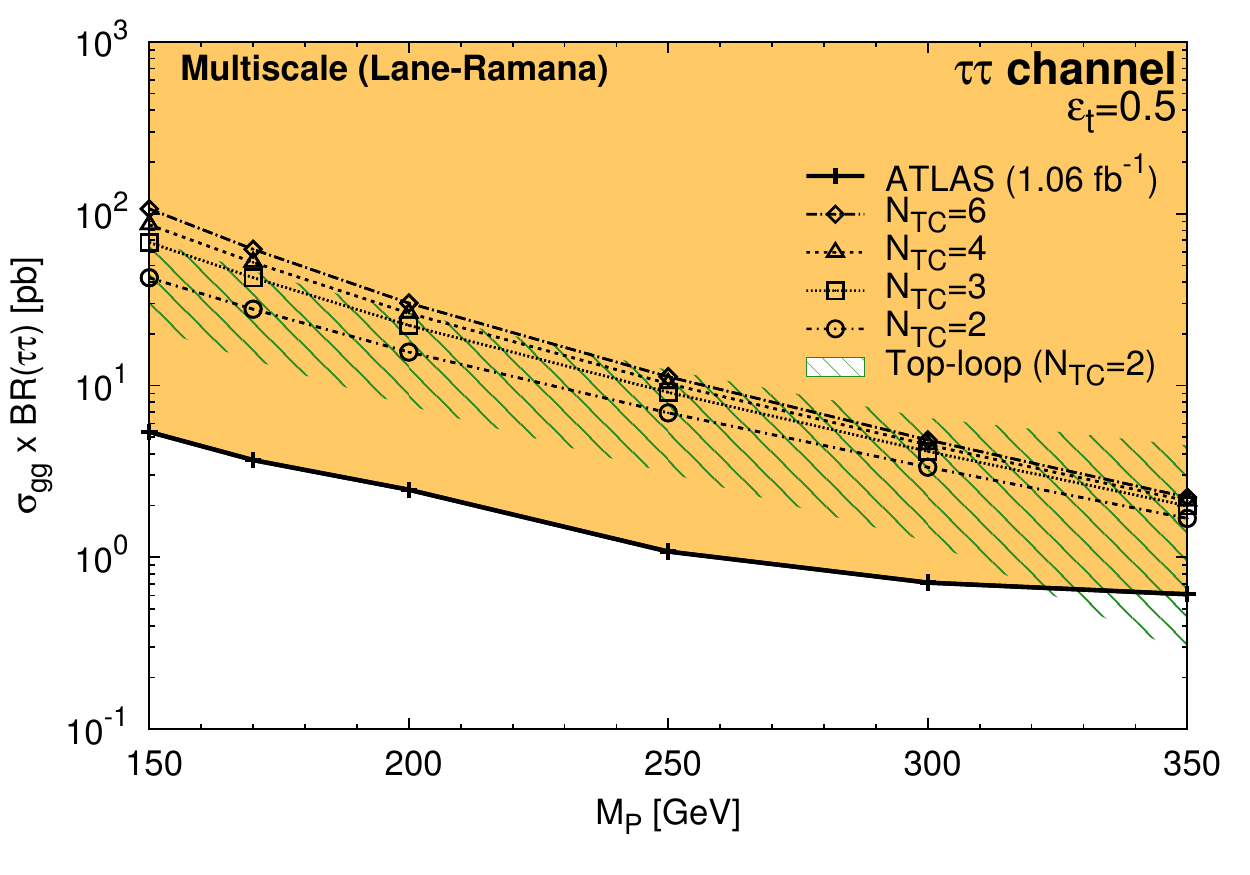}
	}
%
\end{center}
	\caption{Comparison of data and theory for production of a new scalar of mass 150 - 350 GeV that decays to tau lepton pairs; here, technipion production through techniquark loops is potentially modified by including production via top quark loops assuming extended technicolor generates most of the top quark's mass.  In each pane, the shaded region (above the solid line) is excluded by the  $95\%$ CL upper limits on $\sigma_h B_{\tau^+\tau^-}$ from ATLAS  \cite{ATLAS-higgs-di-tau-high}.  As in Figure 4, each pane displays the theoretical prediction (including techniquark loops only) from one technicolor model with colored technifermions, as a function of technipion mass and for several values of $N_{TC}$.  Values of $M_P$ and $N_{TC}$ for a given model that are not excluded by this data are shown as solid (green) symbols.  The hatched region indicates (for $N_{TC} = 2$)  how including the contributions of top-quark loops could impact the model prediction, assuming $\epsilon_t = 0.5$.  If the top and techniquark loop contributions interfere constructively, the model prediction moves to the top of the hatched region; if they interfere destructively, the model prediction moves to the bottom of the hatched region.}
	\label{fig:XSBRggHtautauATLASTau-log-lambda-0-5}
	\end{figure}

First, it means that for technipions heavy enough to decay to top-quark pairs that channel will dominate, so that the branching fractions to $\tau^+\tau^-$ and $\gamma\gamma$ become negligible.  So these models can be constrained by the LHC data discussed in this paper only for $M_P < 2 m_t$.  Second, it implies that charged technipions $P^+$ that are lighter than the top quark can open a new top-quark decay path:  $t \to P^+ b$.   Existing bounds on this decay rate preclude charged technipions lighter than about 160 GeV; for simplicity, we will take this to be an effective lower bound on the mass of our neutral technipions in our discussion here. Finally, as illustrated by the hatched regions in
Fig. \ref{fig:XSBRggHtautauATLASTau-log-lambda-0-5}, top and techniquark loop contributions to technipion production will interfere -- potentially strengthening or weakening the bounds discussed here.

\section{Discussion and Conclusions}
\label{axax}

This first set of LHC data has excluded a large class of technicolor and topcolor-assisted technicolor models that include colored technifermions -- unless the technipions states can be made relatively heavy or the extended technicolor sector can be arranged to cause interference between top-quark and techniquark loops.  Model builders will need to either identify specific technicolor theories able to withstand the limits discussed here,\footnote{For a discussion of possible model-building directions, see \protect\refcite{Chivukula:2011ue}.}while generating the top quark mass without excessive weak isospin violation or FCNC,  or else seek new directions for a dynamical explanation of the origin of mass.   Finally, we would like to stress that additional LHC data that gives greater sensitivity to new scalars decaying to $\tau^+\tau^-$ or that addresses scalars with masses over 145 GeV decaying to $\gamma\gamma$ could quickly probe models down to the minimum number of technicolors and up to higher technipion masses.

\bigskip

\section{Acknowledgments}
The work of RSC and EHS was supported, in part, by the US National Science Foundation under grant PHY-0854889; these authors also gratefully acknowledge the support of the Kobayashi-Maskawa Institute at Nagoya University for travel support to attend this conference. JR is supported by the China Scholarship Council.  PI was supported by the Thailand Development and Promotion of Science and Technology Talents Project
(DPST).


\end{document}